\documentstyle[aps,prl]{revtex}
\begin{document}
\draft
\twocolumn[\hsize\textwidth\columnwidth\hsize\csname @twocolumnfalse\endcsname
\title{Critical dynamics of singlet excitations in a frustrated spin system}
\author{Valeri N. Kotov$^1$, Michael E. Zhitomirsky$^2$, and  Oleg P. Sushkov$^{2,3}$}
\address{
$^1$ Department of Physics, University of Florida, Gainesville, FL 32611-8440, USA\\
$^2$Institute for Theoretical Physics, ETH-H\"onggerberg, CH-8093 Z\"urich,
 Switzerland\\
$^3$School of Physics, University of New South Wales, Sydney 2052,
Australia
}

\maketitle

\begin{abstract}
We construct and analyze a  frustrated quantum spin model with
plaquette order,  in which the
low-energy dynamics is   controlled  by spin
singlets. At a critical value of frustration the
singlet spectrum becomes gapless, indicating a quantum transition
to a phase with dimer order. The magnetic susceptibility  has an
activated form throughout the phase diagram, whereas the specific heat
exhibits a rich structure and a power law dependence on
temperature at the quantum critical point.
This generic behavior can be relevant to quantum antiferromagnets
on Kagom\'e and  pyrochlore lattices where (almost) all
low-energy excitations are spin singlets, as well as to the
 CaV$_4$O$_9$ lattice
 and the
 strongly frustrated antiferromagnets
Cu$_{2}$Te$_{2}$O$_{5}$(Cl,Br)$_{2}$ and
 Li$_{2}$VO(Si,Ge)O$_{4}$.

\end{abstract}

]

It was argued  by Haldane  and
Chakravarty, Halperin and Nelson\cite{HC} that
a quantum phase transition between a N\'eel state and
a magnetically disordered phase
in a two-dimensional (2D) quantum antiferromagnet  is
described by the 2+1 dimensional non-linear $O(3)$ $\sigma$-model.
More recently the problem was  studied in detail using the  $1/N$
expansion\cite{Sach} ($N=3$ is the number of components of
 the order parameter),
numerical methods\cite{Hida} as well as the  Brueckner method
\cite{Kotov,Shevchenko}. In this description  the lowest
quasiparticle excitation is a spin triplet.
There are cases when a  low-energy
 singlet excitation also appears,
however it turns out to be
irrelevant to the critical dynamics because
of its vanishing spectral weight at
the  quantum
critical point\cite{Read}.

The above picture based on  S=1
excitations describes a wide class of quantum
phase transitions \cite{subir}. On the other hand
in Cr-based S=3/2 Kagom\'e materials it
 was found that at low temperatures
($T \lesssim 3-5$~K) the  specific heat $C\propto T^2$ and  is
practically independent of magnetic field up to 12~T \cite{Ramirez}.
If we try to understand these data in terms of
quasiparticles,
the temperature behavior indicates a linear dependence
of the excitation energy on momentum, and the insensitivity to  field
suggests  that the quasiparticle spin is zero.
Finite cluster numerical simulations
for the  S=1/2 Kagom\'e model indeed  show that the lowest excitations are
spin singlets with very small or zero gap\cite{Lhuillier}.
Similar behavior with many singlet states inside the triplet
gap was also  found in the  pyrochlore antiferromagnet \cite{Canals}
 and the  CaV$_4$O$_9$ lattice \cite{Mila}.
In the Kagom\'e case the  singlet quasiparticle
picture is far from being certain. It is unclear how
to explain in this scenario the rather large zero temperature magnetic
susceptibility observed in the Kagom\'e materials \cite{Ramirez}.
It is also possible that the  material is a spin glass and hence a
description in terms of quasiparticles would not be adequate. Nevertheless,
it is very important and interesting to analyze the quasiparticle
scenario. From this point of view the Kagom\'e magnets
are critical (or close to critical) systems, and they
certainly can not be described by the $O(3)$
$\sigma$-model. In the present work we consider a two-dimensional
spin system which exhibits a non-trivial singlet dynamics leading
to a quantum critical point controlled by singlet excitations.

Critical singlet dynamics can naturally appear in a
2D quantum spin model
with a zero temperature quantum phase transition separating two
singlet ground states, both of which are magnetically disordered
but  have different discrete lattice symmetries.
We show that a  spin model
 constructed  as an
array of weakly coupled
frustrated plaquettes  has only singlet
low-energy degrees of
freedom in a certain range of parameters.
Many theoretical works have been  devoted to the
conventional order-disorder quantum phase transition
in similar models,
mainly in relation to the  CaV$_4$O$_9$ compound
\cite{Ueda,Subir,Zheng}.
Different types of singlet dimer states
have also been discussed
for the quantum dimer model\cite{RK},
the Heisenberg
model on a 3D pyrochlore lattice \cite{Harris}, and the generalized
2D  $J_1\!-\!J_2$ model \cite{Rajiv}.
However, low-energy critical singlet dynamics for these systems
has not been studied until now.
Very recent numerical  work has shown, in the context of the 2D
 $J_1\!-\!J_2$ model,
 that a quantum transition governed by singlet dynamics takes place for
  strong frustration $J_{2}\approx J_{1}/2$
  \cite{j1j2theo}. It was also recently argued that this regime is
   experimentally accessible in Cu$_{2}$Te$_{2}$O$_{5}$(Cl,Br)$_{2}$ and
 Li$_{2}$VO(Si,Ge)O$_{4}$
 \cite{j1j2exp}, and measurements of the excitation spectrum of these
 materials, such as Raman spectroscopy,  are now actively being pursued
\cite{peter}.
Let us  mention that dimer states appear also in various 1D spin models
(e.g. in the frustrated $J_1\!-\!J_2$ Heisenberg chain).
Their critical behavior is understood in terms of
deconfined spin-1/2 excitations or spinons, which do not
generally exist in higher dimensions.
A 1D system with features similar to the generic critical
behavior we analyze in this work  is
an effective 1D Kagom\'e lattice  model  \cite{Azaria}
with gapless singlet and gapped triplet excitations.
However, unlike the behavior we focus on,
this model  does not exhibit a  quantum phase transition and  breaking
of lattice symmetry.  Thus our future discussion
is relevant to models in
$D>1$, and a 2D case is analyzed as a generic example.

Consider a cluster of four S=1/2 spins (plaquette, see Fig.1(a)):
\begin{equation}
\label{hpl}
\hat{\cal H}_0=J_1\sum_{\langle ij\rangle}{\bf S}_i\cdot {\bf S}_j
+J_2\sum_{(ij)}{\bf S}_i\cdot {\bf S}_j,
\end{equation}
where $\langle ij\rangle$  and
$(ij)$ denote, respectively,  the side and diagonal bonds.
We choose the diagonal coupling $J_2$ to be close
to the side one: $J_2\approx J_1=J> 0$, and define
 $\alpha= J_1-J_2 \ll J$.
In this regime the  spectrum of a  plaquette is the  following:
 two close singlet
states which we denote by $|s_A\rangle$ and $|s_B\rangle$, with energy
difference
$\epsilon_B-\epsilon_A = 2\alpha$,
three almost degenerate triplet states with excitation energy $J$
above the singlets, and one S=2 state with  energy $3J$.
The two singlet wave functions are expressed as:
$|s_A\rangle=\frac{1}{\sqrt{3}}\{[1,2][3,4]+[2,3][4,1]\}$
and $|s_B\rangle=\{[1,2][3,4]-[2,3][4,1]\}$, where $[i,j]$ denotes
a singlet formed by  the nearest-neighbor spins $i$ and $j$ (Fig.1(a)).
Both  singlets  $|s_A\rangle$ and $|s_B\rangle$ are invariant
under a four-fold rotation of
a plaquette,
whereas the columnar dimer states, e.g.\ $[1,2][3,4]$,
break this symmetry.
Consider now two decoupled plaquettes which for $\alpha=0$
have a $2\times 2$ degenerate ground state. The degeneracy
is lifted if a weak interaction between the plaquettes is switched on
selecting one of the columnar
dimer states. Such a degeneracy lifting is different from
the effect of a non-zero $\alpha$ and competition between the two leads to a nontrivial
singlet dynamics in an array of weakly coupled plaquettes.

To be specific consider a square array with
antiferromagnetic couplings
$j_1$ between nearest-neighbor
spins  and $j_2$ between next-nearest-neighbor (diagonal) spins
from different plaquettes, see Fig.1(b).
For  $j_1=J_1$ and $j_2=J_2$ it would
be equivalent to the translationally-invariant
 2D $J_1-J_2$ model.
However now we are interested in the  weak-coupling limit:
$j_1,j_2 \ll J_2\approx J_1=J$. The low-energy singlet sector
of the Hilbert space has a natural pseudospin representation
in terms of the  states $|\!\!\uparrow\rangle=|s_A\rangle$ and
$|\!\!\downarrow\rangle=|s_B\rangle$. The total Heisenberg spin Hamiltonian
of the array of coupled plaquettes is
mapped on the following pseudospin Hamiltonian to lowest order
in the small parameters $\alpha$ and $j_{1,2}/J$:
\begin{eqnarray}
\hat{\cal H}&=& -\Omega \sum_{\langle i,j\rangle}
\Bigl[\case{1}{6} S^z_i S^z_j + \case{1}{2}S^x_iS^x_j
  + \case{e^{i{\bf Q}({\bf i}-{\bf j})}}
{2\sqrt{3}} (S^z_iS^x_j \!+ \!S^x_iS^z_j)\Bigr] \nonumber \\
 & &  -\Omega h \sum_i S^z_i \ .
\label{pseudospin}
\end{eqnarray}
The energy scale $\Omega$ and the effective dimensionless
``magnetic field'' $h$ are obtained in second-order perturbation
 theory:

\begin{equation}
\Omega= \frac{1}{2J}(j_1^2+j_2^2), \ h = \frac{1}{\Omega}
[2\alpha+(j_1^2+j_2^2-6j_1j_2)/6J].
\end{equation}
 We have also defined
${\bf Q} = (0,\pi)$ and   dropped a
constant term. The first sum in Eq.(\ref{pseudospin}) is
over  nearest-neighbor plaquettes since bonds
connecting second neighbors
contribute to a constant term only and thus do not change
the singlet dynamics.
 We stress that (\ref{pseudospin}) is an exact mapping
of the original Heisenberg model  in the low-energy sector
(excitation energy $\omega \sim \Omega , \alpha \ll J$).
The Hamiltonian $\hat{\cal H}$ describes an anisotropic
ferromagnet in an external field. The interaction between pseudospins
on adjacent sites is purely Ising, which can be seen
by rotating the spin axes. However the Ising axis
is staggered in the $x$--$z$ plane deviating by
angle $\pm \pi/6$ from the $x$-direction
for horizontal (vertical) pairs.

The form of the Hamiltonian Eq.(2) is unaffected by asymmetry in
the coupling constants $j_1$ and $j_2$, which  would only change the position
of the reference zero-field level. In fact, the pseudospin
Hamiltonian remains the same with
 parameters given by:
$\Omega=j_1^2/2J$ and $h=[2\alpha+(j_1^2-3j_1j_2)/6J]/\Omega$,
even if we switch off one of
the diagonal $j_2$-bonds between nearest-neighbor plaquettes.
In this case our model resembles the frustrated $\case{1}{5}$-depleted
square lattice of CaV$_4$O$_9$ \cite{Ueda,Subir,Zheng},
which has next-nearest-neighbor couplings comparable or even
exceeding the nearest-neighbor exchange.
Thus our results
 apply  to the  singlet ground states
of this magnet as well.

In zero ``magnetic field'' $h=0$ the ground state of $\hat{\cal H}$
has  broken Ising symmetry with
all spins parallel or antiparallel to the $\hat{x}$
direction. In the language of spin singlets  breaking of Z$_2$
symmetry
corresponds to  spontaneous dimerization in one of
the two columnar patterns. The rotational
symmetry of the square lattice is obviously broken.
The ``magnetic field''  tends to orient all spins
along the $\hat{z}$ axis,  and  depending on the sign
of $h$ favors either $|s_A\rangle$ or $|s_B\rangle$.
Hence there are two critical fields, a positive and a negative one,
 for  transitions
into states with a restored Z$_2$ symmetry.
Since Eq.(2)
is  invariant under $h\rightarrow-h$, we consider
the case $h>0$ only.

To study in more detail the properties of the  symmetric
phase near the critical point we map the pseudospin
Hamiltonian (\ref{pseudospin}) onto a
hard-core boson model. Positive and large $h$ corresponds to
$J_1>J_2$. In
this case we choose the bare ground state as
$|0\rangle =\prod_i|s_A\rangle_i$.
A boson creation operator on  site $i$ is defined by
$|s_B\rangle_i = b_i^\dag|s_A\rangle_i$. In terms of pseudospins we have
  $S^z_i=\case{1}{2}-b^\dag_ib_i$, $S^-_i=b^\dag_i$.
The resulting  boson Hamiltonian is:
\begin{eqnarray}
\label{heff1}
&&H_B=\epsilon\sum_ib_i^{\dag}b_i+t\sum_{\langle ij\rangle}
(b_i^{\dag}b_j+
b_i^{\dag}b_j^{\dag}+h.c.)+\\
&&g\sum_{\langle ij\rangle}
\pm (b_i^{\dag}b_j^{\dag}b_i+b_i^{\dag}b_j^{\dag}b_j+h. c. )+
V\sum_{\langle ij\rangle} b_i^{\dag}b_j^{\dag}b_jb_i. \nonumber
\end{eqnarray}
In the $g$-term the sign $+$
corresponds to a horizontal link and the sign $-$ to a vertical one.
The parameters are:
\begin{equation}
\epsilon=\Omega (\case{1}{3}+h)\ , \ \
t= -\case{\Omega}{8}\ , \ \
g=-\case{\Omega}{4\sqrt{3}}\ , \ \
V=-\case{\Omega}{6}\ .
\label{par}
\end{equation}
Considered in combination with  the hard-core constraint $(b_i^{\dag})^2=0$,
 the
Hamiltonian (\ref{heff1}) is equivalent to
the pseudospin Hamiltonian (\ref{pseudospin}) and hence it is an exact
mapping of the  original spin problem  in its low-energy sector.

The bosonic form of the effective Hamiltonian is convenient for
the analysis of
 the low density (disordered) phase.
Diagonalization of Eq.(4) in the quadratic approximation gives the
following excitation spectrum: $\omega_{\bf k}= \sqrt{(\epsilon+4
t\gamma_{\bf k})^2-(4t\gamma_{\bf k})^2}$, where $\gamma_{\bf k}=
\case{1}{2}(\cos k_x +\cos k_y)$. We set the  inter-plaquette
lattice spacing to one. The excitation gap $\Delta=\omega_{{\bf
k}=0}$ vanishes at
 the critical point $h_c= \case{2}{3}$.
The other critical point is at $-h_c$.
For $h>h_c$ the model is in  the A-type
plaquette phase, for $h<-h_c$ it is in the B-type plaquette
phase, and in between $-h_c<h<h_c$ there is
a phase with a boson condensate at ${\bf k}=0$.
The condensate is doubly degenerate because the boson
field can have both signs.

The zero-point quantum fluctuations, which exist in the
disordered (plaquette) phases, change
slightly the critical field $h_c$.
The simplest way to account for the  correlation effects is to use the Brueckner
technique developed in Ref.~\cite{Kotov}.
The cubic and quartic interactions in
(\ref{heff1}) are treated in the one-loop approximation,
and the hard-core constraint is enforced via an infinite repulsion term
$U\sum_i b_i^{\dag}b_i^{\dag}b_ib_i$, $U \to \infty$,
added to the Hamiltonian (\ref{heff1}).
This interaction is taken into account
 in the  Brueckner (low-density gas) approximation.
The singlet gap obtained from the self-consistent solution of
the resulting  Dyson equations
is plotted in Fig.2. At the critical point the  small parameter which
justifies the Brueckner approximation is the singlet
density  $\langle b_i^{\dag}b_i\rangle
\approx 0.04$.
The renormalized critical value $h_c\approx 0.65$ is only slightly
below the result for non-interacting magnons.
The reason is the strong compensation
 of the hard-core corrections by the
one-loop diagrams arising from the cubic terms in Eq.(4).
 The dependence of the gap on $h$
near the transition point agrees with the critical index $\nu=0.63$, expected
for the $O(N=1)$  $\sigma$ model \cite{NL} and is quite different from the
 value
$\nu = 0.5$ for non-interacting magnons.

 Next we analyze the singlet excitation spectrum
in the ordered dimer phase, $h<h_c$. We return  to the
pseudospin representation (\ref{pseudospin}) and use the analogy
between the  evolution of the dimer phase and the
spin reorientation process in an external magnetic field.
The  Holstein-Primakoff transformation is applied
to the pseudospins in the
rotating coordinate frame, which to lowest order
 coincides with the  hard-core boson representation.
At $h=0$ all spins point along the $\hat{x}$-axis.
For finite $h$ the spins tilt towards the field direction
at an angle $\sin\theta=h/h_c$. Keeping only quadratic
terms we find the following ``classical'' spectrum of singlet
excitations:
\begin{equation}
\omega_{\bf k} = \Omega \sqrt{1-\gamma_{\bf k}
(1-\case{2}{3}\cos^2\theta)+ \case{1}{\sqrt{3}}\sin 2\theta
\gamma_{\bf k+Q}} \ . \label{wk}
\end{equation}
In accordance with the broken discrete symmetry the
singlet excitations have a finite gap $\Delta = \omega_{{\bf k}=0}=
\Omega\sqrt{2(1-h^2/h_c^2)/3}$, plotted in Fig.2. The gap vanishes at
the critical point.
Proceeding further with the spin analogy we have also calculated the
``spin reduction'' for the ferromagnet (\ref{pseudospin}).
This parameter shows how far is the real ground state
wave-function from the
approximate mean-field ansatz. At $h=0$ we find
$\langle S\rangle\approx 0.498$, meaning that the effect of
 quantum fluctuations
is extremely small and the linear spin-wave theory is well justified.
At the transition point $h=h_c$ the zero-point oscillations are larger
$\langle S\rangle\approx 0.44$
but still small enough to justify our approximations
leading to  Eq.(\ref{wk}).

The broken Z$_2$ symmetry in the ground state
at $|h|<h_c$ leads to a finite temperature transition.
The whole phase diagram of the system of coupled plaquettes in
the $h$--$T$ plane is shown schematically in Fig.3.
The critical temperature can be estimated
as $T_c(h=0)\simeq \Omega$ ($1.14\Omega$ for a pure Ising case).
The phase
transition between the ordered dimer and the plaquette states
belongs to the 2D Ising
universality class, and we  therefore expect a logarithmic
singularity in the specific heat:
$C\sim \ln|T-T_c(h)|$.
Below $T_c$ the excitations
acquire a gap and $C$ goes exponentially at
low temperatures. The same activated dependence
holds also for the symmetric phases at $|h|>h_c$.
However, when we approach the quantum critical points at $h=\pm h_c$
the gap becomes smaller and the low-$T$
behavior of the specific heat changes to the quantum critical
law $C_V =\gamma T^2$.  The  universal prefactor
$\gamma$ can be calculated  using the Brueckner technique
\cite{Shevchenko}. The  result is
$\gamma={{12\zeta(3)}\over{5\pi c^2}} \approx 0.92/c^{2}$,
per plaquette,
where $c=\Omega/2$ is the singlet excitation velocity at the
critical point (we set
$k_B=\hbar=1$). This value coincides with the large-N mean-field
 result \cite{Sach}. Notice that, unlike $C_V$,
the magnetic susceptibility is always activated
$\chi \sim {\mbox{exp}}(-\Delta_t/T), T \rightarrow 0$, since
 it is governed by the triplet gap $\Delta_t \sim J$.

The variation of $C_V$ as a function of temperature
is schematically presented in Fig.4 for the different parts of
the phase diagram. In addition to the singlet contribution, whose
 form was discussed above, we have also shown  the peaks expected to arise from
 the higher energy triplet and quintiplet states.
 We remind the reader
that $\Omega$ and $J$ represent two distinctly different energy scales
 since according to our weak inter-plaquette coupling assumption
 $\Omega \ll J$.
We have estimated that  the singlets contribute about 25\% to the
 entropy $S=\int (C_V/T) dT$ (and  61\% and 14\% are taken by
the S=1 and S=2 states,
 respectively).
 The possibility of rich
behavior at low T has been debated for
some time for the Kagom\'e antiferromagnet  and a sharp structure
 is believed to exist due to low-energy singlets
\cite{Sindzingre}. Notice that in our model
 a very peculiar behavior with a logarithmic singularity  sets in at
 $h < h_c$ which  crosses over to the quantum critical regime at
$h=h_c$. Such a sharp feature is  generic for systems near
 a transition between two singlet ground states.
 We also note that near the quantum critical point
 $h \approx h_{c}$,  $T_{c}$ is expected to be small
 (and ultimately vanish at $h=h_{c}$). Consequently
  the logarithmic singularity at $T=T_{c} \ll \Omega$ and
 the singlet peak at $T \sim \Omega$ should be clearly  separated
 (this has been assumed in Fig.4). Alternatively, for $h \ll 1$
 one has $T_{c} \sim \Omega$ and therefore the two contributions should merge.
 A numerically reliable calculation of $C_{V}$ in the temperature
 region $T<\Omega$ is a separate problem, however for higher temperatures (of order $J$) a
 quantitative description can be easily achieved, and the result  is plotted in
  Fig.5 (for the  specific value of $\Omega=0.1J$).
The contributions of the triplet and quintiplet states have merged into
 a broad peak around $T \approx J/3$, while the singlet peak is sharper and
 centered at $T \approx \Omega/2$ (in this region our calculation is expected
 to be qualitatively correct).

In conclusion, we have analyzed a quantum spin model with purely
 singlet low-energy dynamics.
There is a quantum phase transition
in the model separating a  disordered plaquette phase and a
columnar dimer phase.
 Even though the inter-plaquette interaction was assumed to be weak, we expect
 our results to
hold also for stronger
interactions as long as there are no other instabilities,
and to be applicable, e.g.  to  CaV$_4$O$_9$.
The broken symmetries in the singlet ground state
of CaV$_4$O$_9$ were considered previously in Ref.~\cite{Subir}
in the framework of the quantum dimer model, and the possibility of
 Ising transitions between spin-Peierls and other disordered phases was
 discussed \cite{comment}.
Furthermore,  recent numerical studies of the
 square-lattice $J_{1}-J_{2}$ model \cite{j1j2theo}
 have indeed found a quantum transition near $J_{2} \approx J_{1}/2$ of the type
 discussed in the present
 work, even though a  formally small expansion parameter (justifying the
 separation of the singlet and triplet dynamics) is not available.

We have found that while  the magnetic susceptibility  is always activated,
 the specific heat  behavior is  very rich and changes substantially  in the  different
regimes.
The model  was inspired  by experimental and numerical data for
Kagom\'e systems and also
 shares  common structure with pyrochlore antiferromagnets,
the  CaV$_4$O$_9$ lattice,  and  strongly frustrated square lattice
antiferromagnets.
The novel quantum critical behavior associated with singlet criticality
 (or proximity to such a critical point)
 discussed in this work
can be relevant to a wide class of disordered quantum spin systems, and
could be detected by measurements of the specific heat as well as
 the low-T excitation spectrum in the
 S=0 sector via Raman spectroscopy.

We are grateful to C. Lhuillier, F. Mila, A.P. Ramirez,  R.R.P. Singh,
S. Sachdev, W.H. Zheng, and P. Lemmens for
stimulating discussions. This work was supported by
NSF Grant
 DMR9357474 (V.N.K.), in  part
 by NSF Grant PHY94-07194 (O.P.S.), and by the Swiss National Fund
 (M.E.Z.).

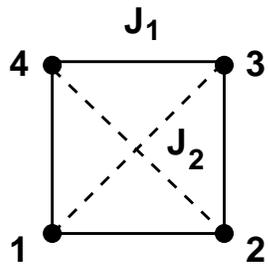
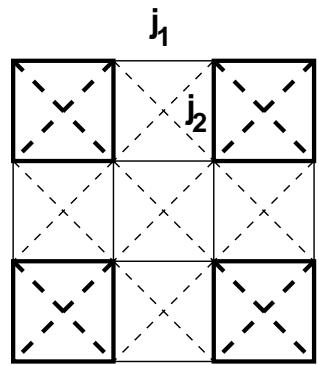
\begin{figure}
\caption
{(a) A single  frustrated plaquette.
(b) An array of interacting plaquettes.}
\label{fig.1}
\end{figure}

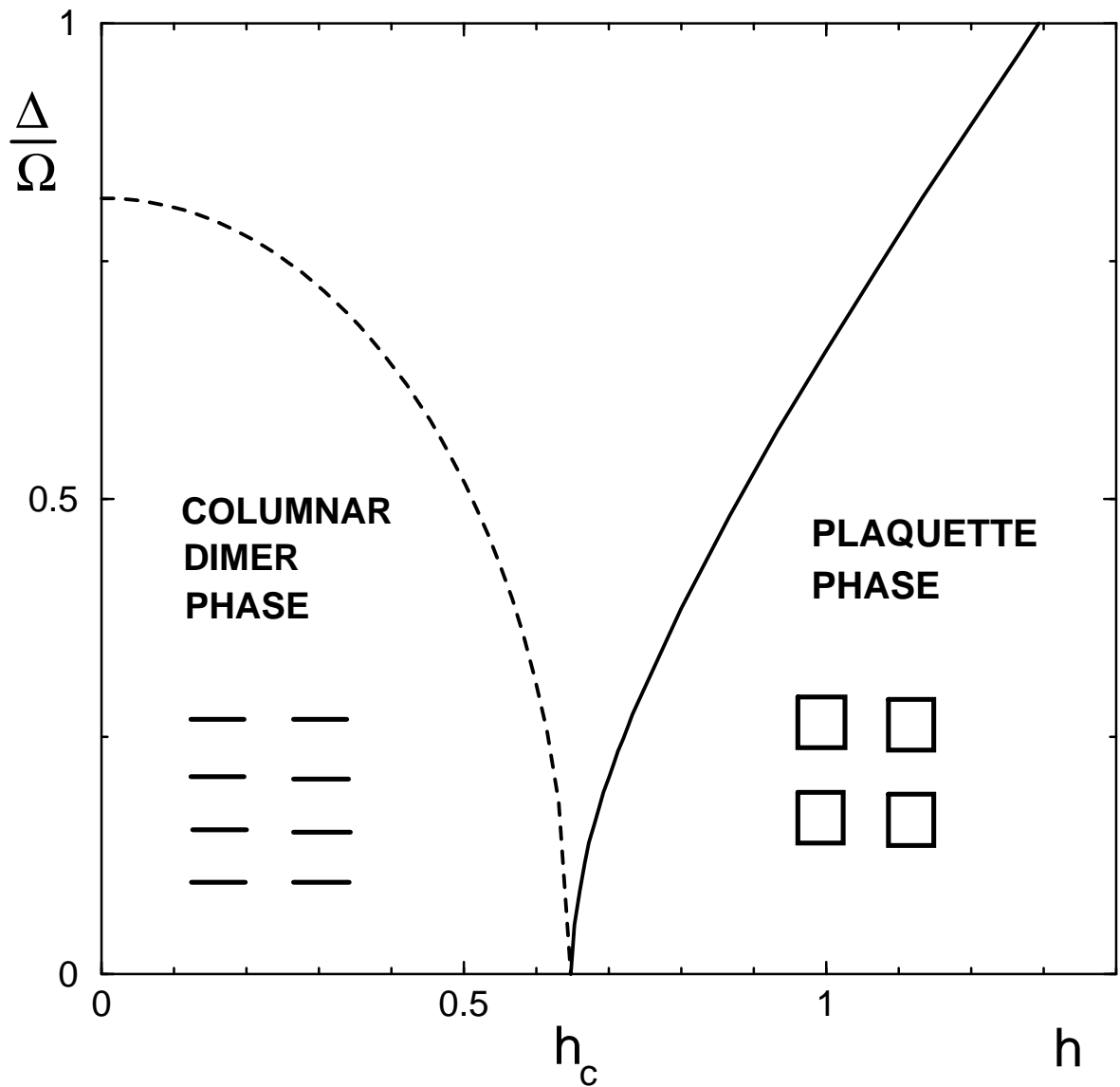
\begin{figure}
\caption
{Singlet gap in the vicinity of the quantum critical point
 separating the plaquette and dimerized phases.
}
\label{fig.2}
\end{figure}

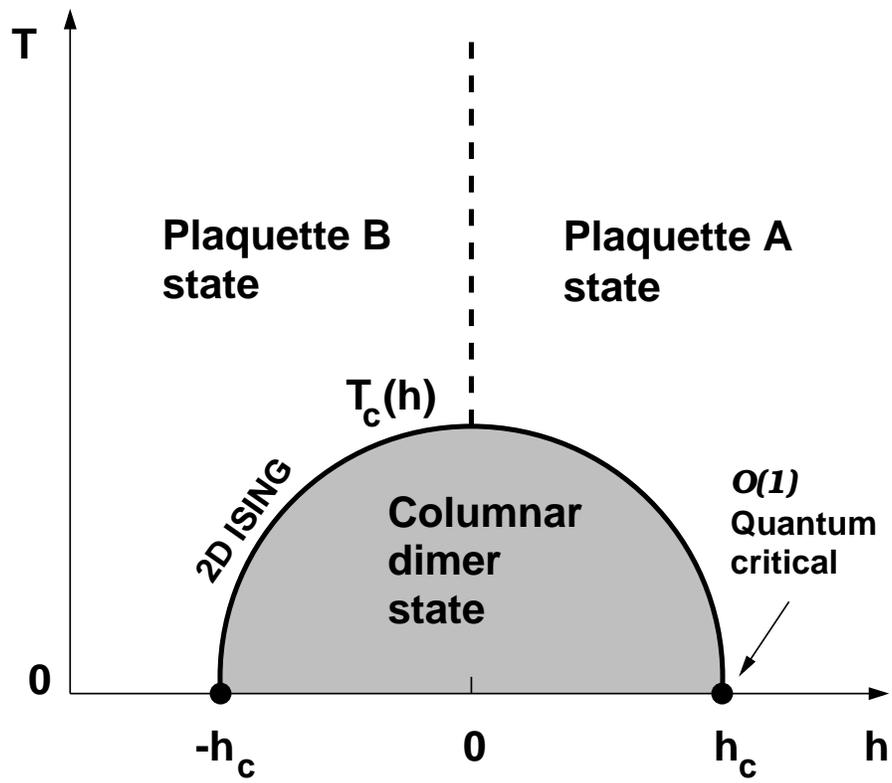
\begin{figure}
\caption
{Schematic phase diagram of the model. The solid line represents
 the 2D Ising critical line, and the dots are the
two quantum critical points  belonging to the $O(1)$
 (equivalently 3D Ising) universality class. The vertical dashed line
 represents the crossover boundary between the two plaquette phases.
}
\label{fig.3}
\end{figure}

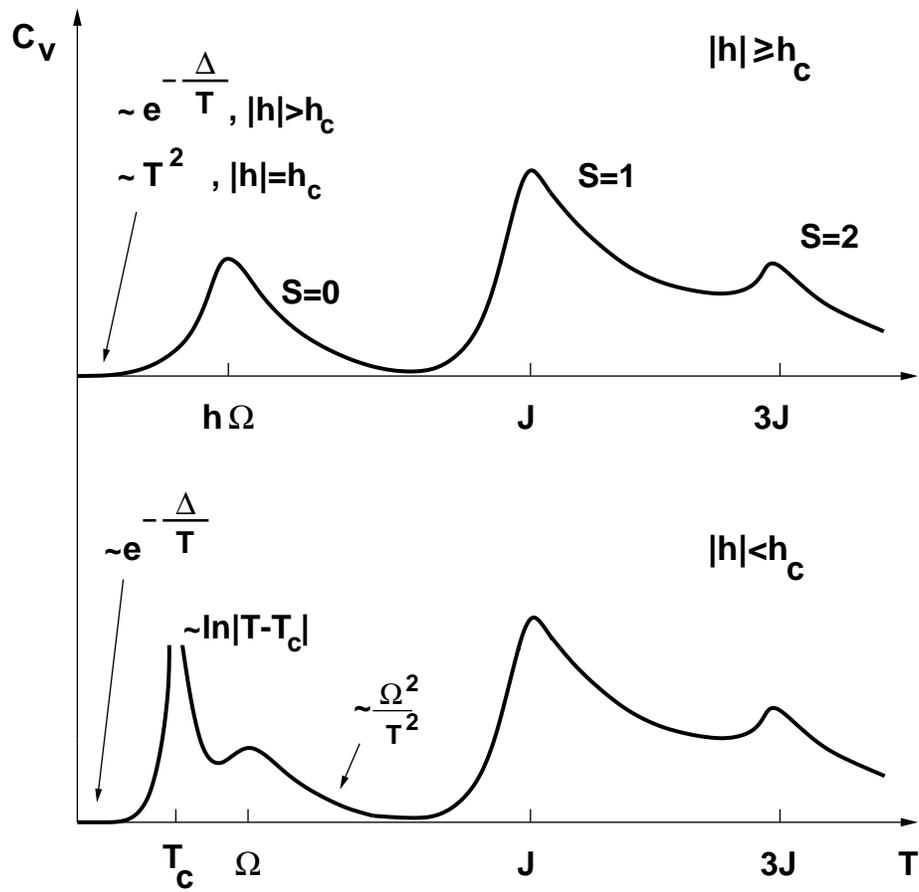
\begin{figure}
\caption
{Schematic temperature dependence of the specific heat throughout the
 plaquette phase, $h>h_{c}$, down to the quantum critical point
$h=\pm h_{c}$ (top),
 as well as  for $h<h_{c}$, across the 2D Ising critical line (bottom).
}
\label{fig.4}
\end{figure}

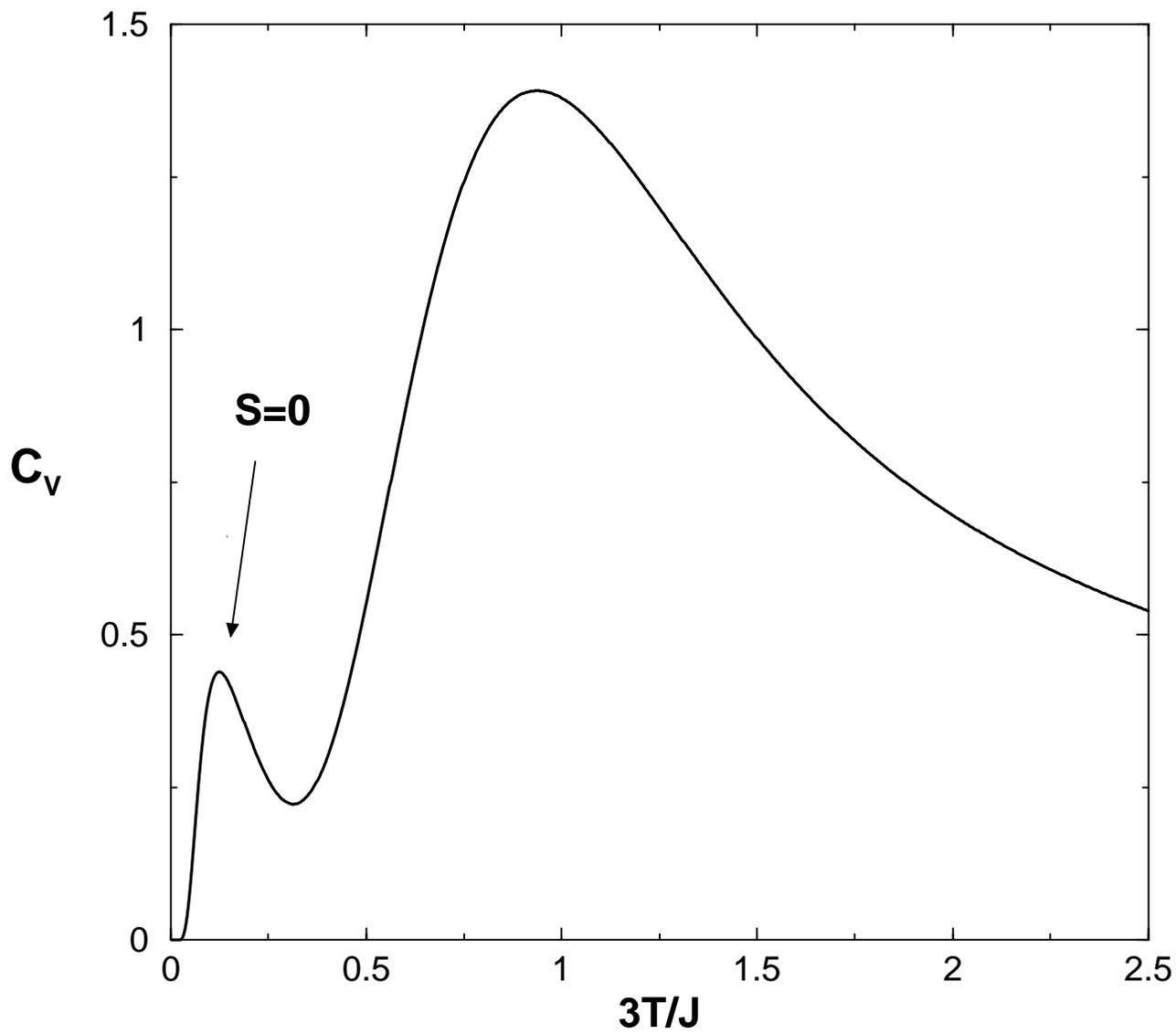
\begin{figure}
\caption
{Calculated specific heat for $\Omega=J/10$,  showing the low-temperature
singlet contribution at $T \approx \Omega/2$ as well as the broader peak
 on the scale of $J$ arising from the $S=1,2$ states.
}
\label{fig.5}
\end{figure}


\begin{references}

\bibitem{HC}
 F.D.M. Haldane, Phys. Rev. Lett. {\bf 50}, 1153 (1983);
S. Chakravarty, B.I. Halperin, and D.R. Nelson,
 Phys. Rev. B {\bf 39}, 2344 (1989).

\bibitem{Sach}
A.V. Chubukov, S. Sachdev, and J. Ye, Phys. Rev. B {\bf 49}, 11919 (1994).

\bibitem{Hida}
A.W. Sandvik and D.J. Scalapino, Phys. Rev. Lett. {\bf 72}, 2777 (1994);
W.H. Zheng, Phys. Rev. B {\bf 55}, 12267 (1997).

\bibitem{Kotov} V.N. Kotov, O.P. Sushkov, W.H. Zheng, and J.
Oitmaa, Phys. Rev. Lett. {\bf 80}, 5790 (1998).

\bibitem{Shevchenko} P.V. Shevchenko, A.W. Sandvik,
and O.P. Sushkov, Phys. Rev. B {\bf 61}, 3475 (2000).

\bibitem{Read} V.N. Kotov and O.P. Sushkov,
Phys. Rev. B {\bf 61}, 11820 (2000).

\bibitem{subir} S. Sachdev, {\it Quantum Phase Transitions}
(Cambridge University Press, Cambridge, 1999).

\bibitem{Ramirez}
 A.P. Ramirez, B. Hessen, and M. Winklemann,
Phys. Rev. Lett. {\bf 84}, 2957 (2000).

\bibitem{Lhuillier} Ch. Waldtmann, H.-U. Everts, B. Bernu, P. Sindzingre,
C. Lhuillier, P. Lecheminant, and L. Pierre, Eur. Phys. J.
 B {\bf 2}, 501 (1998).

\bibitem{Canals} B. Canals and C. Lacroix,
Phys. Rev. Lett. {\bf 80}, 2933 (1998).

\bibitem{Mila} M. Albrecht, F. Mila, and D. Poilblanc,
Phys. Rev. B {\bf 54}, 15856 (1996).

\bibitem{Ueda}
K. Ueda, H. Kontani, M. Sigrist, and P.A. Lee, Phys. Rev. Lett.
{\bf 76}, 1932 (1996); O.A. Starykh, M.E. Zhitomirsky, D.I.
Khomskii, R.R.P. Singh, and K. Ueda, {\it ibid.} {\bf 77}, 2558
(1996).

\bibitem{Subir} S. Sachdev and N. Read, Phys. Rev. Lett. {\bf 77}, 4800 (1996).

\bibitem{Zheng} W.H.  Zheng, M.P. Gelfand, R.R.P. Singh, J.
Oitmaa, and C.J. Hamer,
 Phys. Rev. B {\bf 55}, 11377 (1997).

\bibitem{RK}
 P.W. Leung, K.C. Chiu and K.J. Runge, Phys. Rev. B {\bf 54},
12938 (1996).

\bibitem{Harris} A.B. Harris, A.J. Berlinsky and C. Bruder,
J. Appl. Phys. {\bf 69}, 5200 (1991); M. Isoda and S. Mori,
 J. Phys. Soc. Jpn.  {\bf 67}, 4022 (1998).

\bibitem{Rajiv} R.R.P. Singh, W.H. Zheng, C.J. Hamer, and J. Oitmaa, Phys. Rev. B {\bf 60}, 7278 (1999).

\bibitem{j1j2theo}
 O.P. Sushkov, J. Oitmaa, and W.H. Zheng, cond-mat/0007329, and references
 therein.

\bibitem{j1j2exp} M. Johnsson, K.W. Tornroos, F. Mila, and P. Millet, submitted to Chem. Mater. (2000);
R. Melzi, P. Carretta, A. Lascialfari, M. Mambrini, M. Troyer, P.
Millet, and F. Mila, Phys. Rev. Lett. {\bf 85}, 1318 (2000).

\bibitem{peter} P. Lemmens, private communication.

\bibitem{Azaria} P. Azaria, C. Hooley, P. Lecheminant, C. Lhuillier, and A. M. Tsvelik,
 Phys. Rev. Lett. {\bf 81}, 1694 (1998).


\bibitem{NL} R. Guida and J. Zinn-Justin, J. Phys. A {\bf 31}, 8103 (1998).

\bibitem{Sindzingre} P. Sindzingre, G. Misguich, C. Lhuillier, B. Bernu, L. Pierre,
 Ch. Waldtmann, and H.-U. Everts,  Phys. Rev. Lett. {\bf 84},
2953 (2000).

\bibitem{comment}
However the  columnar dimer  ordering  found in the present work
never corresponds to the ground state in \cite{Subir},  possibly
indicating a weakness of the quantum dimer model \cite{Subir}
 applied to a real spin system with Heisenberg
interactions.

\end{references}
\end{document}